\documentclass[aps,twocolumn,pra,showpacs,superscriptaddress,floatfix]{revtex4-1}
\usepackage{latexsym,amsmath,amssymb,amsfonts,graphicx,color,amsthm}
\usepackage{enumerate}
\usepackage{braket}
\usepackage{adjustbox}
\usepackage{footnote}
\usepackage[dvipsnames]{xcolor}
\usepackage{tipa}
\usepackage{textcomp}
\usepackage{fontenc}
\usepackage{mathrsfs}
\usepackage{colortbl}
\usepackage{txfonts}
\usepackage{graphics,graphicx}
\usepackage{mathtools}
\usepackage{epsfig}
\usepackage{epstopdf}
\usepackage{dsfont}
\usepackage[colorlinks=true,linkcolor=blue,urlcolor=magenta,citecolor=red]{hyperref}
\usepackage[normalem]{ulem}
\usepackage[caption=false,labelformat=simple]{subfig}

\begin{document}


\title{Bright and polarized fiber in-line single photon source based on plasmon-enhanced emission into nanofiber guided modes}

\author{K. Muhammed Shafi}

\affiliation{Center for Photonic Innovations and Institute for Laser Science, University of Electro-Communications, Tokyo 182-8585, Japan.}
\affiliation{Deltafiber Ltd., Yaehara, Tomi-shi, Nagano 389-0406, Japan.}  
\affiliation{Department of Instrumentation and Applied Physics, Indian Institute of Science, Bengaluru 560012, India.}
\author{Ramachandrarao Yalla}
\affiliation{School of Physics, University of Hyderabad, Hyderabad, Telangana 500046, India.}
\author{Kali P. Nayak}
\email{kalipnayak@uec.ac.jp}
\affiliation{Center for Photonic Innovations and Institute for Laser Science, University of Electro-Communications, Tokyo 182-8585, Japan.}
\affiliation{Department of Engineering Science, University of Electro-Communications, Tokyo 182-8585, Japan.}


\begin{abstract}
We demonstrate a bright and polarized fiber in-line single photon source based on plasmon-enhanced emission of colloidal single quantum dots into an optical nanofiber. We show that emission properties of single quantum dots can be strongly enhanced in the presence of single gold nanorods leading to a bright and strongly polarized single photon emission. The single photons are efficiently coupled to guided modes of the nanofiber and eventually to a single mode optical fiber. The brightness (fiber-coupled photon count rate) of the single photon source is estimated to be $12.2\pm0.6$ MHz, with high single photon purity ($g^2(0)=0.20\pm0.04$) and degree of polarization as high as 94-97\%. The present device can be integrated into fiber networks paving the way for potential applications in quantum networks.
\end{abstract}

\maketitle

\section{Introduction}
Development of efficient single photon sources (SPS) is central to the realization of quantum networks and quantum information science\,\cite{Kimble, OBrien, Arakawa}. A straightforward approach to generate single photons is to collect the emission from a single quantum emitter \cite{Aharonovich,Senellart}. However, there are two key challenges in this approach. One is the emission properties of the quantum emitter and the other is efficient collection of emitted single photons in a single spatial mode preferably into a single mode optical fiber (SMF).

The requirements as a quantum emitter for an efficient SPS is that it must be bright, producing a high photon count rate and with a high degree of polarization (DOP). High photon count rate is essential to match the requirements for high data rates. High DOP is also crucial as many protocols require information to be encoded as polarization qubits \cite{Bennett,Nielsen,Nicolas}. In this context, solid-state quantum emitters like quantum dots (QDs) or atom-like defects in crystalline hosts \cite{Arakawa,Aharonovich,Senellart,Michler} are one of the promising choices for practical SPS, based on the emission properties brighter than neutral atoms or ions and easier techniques to isolate single emitters. However, the presence of the complex mesoscopic solid-state host induces various challenges, e.g. inhomogeneous spectral broadening, emission intermittency and reduced DOP, etc. In the last decade, there have been significant efforts and developments to overcome such issues \cite{Aharonovich,Senellart,Park,Lin,Senellart2}. 

On the other hand, efficient collection of emitted single photons is an inevitable requirement for high data rates and deterministic communication protocols. In this direction, nanophotonic waveguides and resonators have shown promising advances taking advantage of the strong transverse confinement of photonic modes beyond free-space optics \cite{Aharonovich,Senellart,Sipahigil,Daveau,Waks}. In particular, metal nanostructures have shown state-of-the-art field confinement leading to strong Purcell enhancement and polarized single photon emission due to their localized surface plasmon resonance (LSPR) \cite{Mulvaney,Esteban,Hoang,Gleb,TongGNR,Blinking2022}. However, the propagation losses in metal waveguides are detrimental. On the other hand, dielectric waveguides have limited field confinement but excellent propagation properties for long-distance communication. 

In this context, tapered optical fiber with subwavelength diameter waist, optical nanofiber (ONF), provides a unique fiber in-line platform for collecting single emitter fluorescence \cite{LTong1,famsan1,Sile1,Rolston,Kali1,LTong2}. The distinct point of the technique is that the guided mode of the ONF can adiabatically evolve to the SMF mode with near-unity efficiency. This provides an automatic and alignment-free fiber-coupled platform for an SPS and easy integration into fiber networks. 
In the last decade, there has been significant development to interface the ONF platform with various single emitters \cite{Kali2,Yalla1,ONFNDarno,FujiwaraQD,Skoff,Schell2,aoki,Kali3}. 
Recently, we have reported a fiber in-line SPS based on a hybrid system of colloidal single QDs deposited on an ONF and cooled down to cryogenic temperature (3.7 K). However, the maximum photon count rate and the decay time were limited to 1.6$\pm$0.2 MHz and 10.0$\pm$0.5 ns, along with limited DOP \cite{ShafiSPS}.

Here, we demonstrate a hybrid quantum system by combining a single QD and a single gold nanorod (GNR) coupled to an ONF. We show that emission properties of single QDs can be strongly enhanced in the presence of GNR leading to a bright and strongly polarized single photon emission. The single photons are efficiently coupled to guided modes of the ONF and eventually to a SMF. The brightness (fiber-coupled photon count rate) of the SPS is estimated to be $12.2\pm0.6$ MHz, with high single photon purity ($g^2(0)=0.20\pm0.04$) and DOP as high as 94-97\%.

\begin{figure}[!ht]

\centering

 \includegraphics[width=\columnwidth]{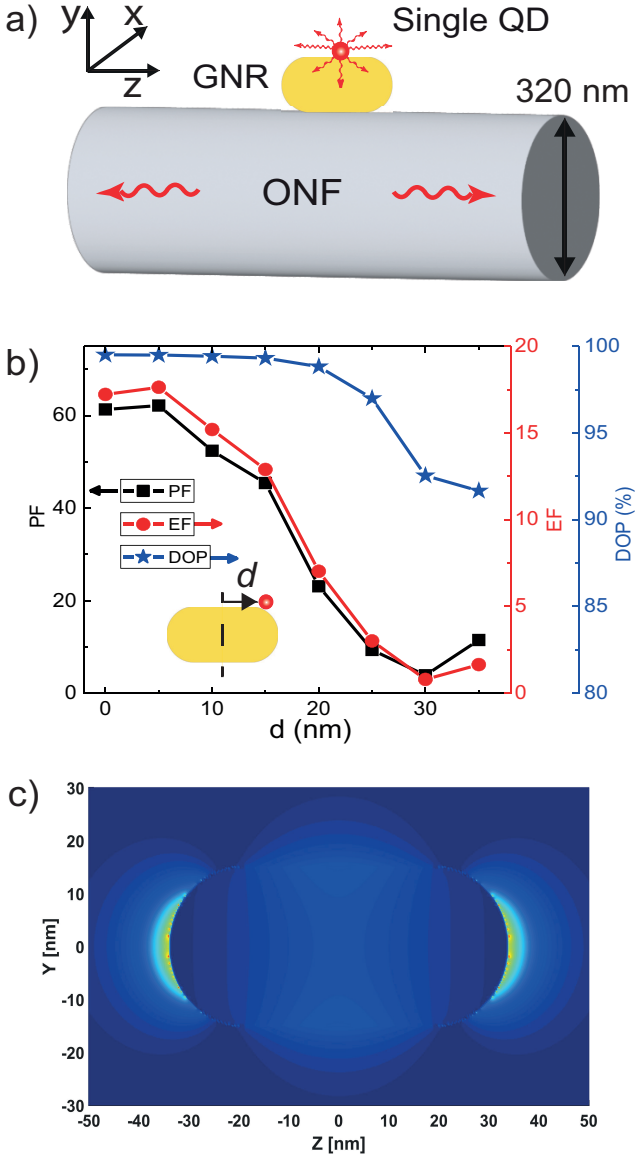}

\caption{Numerical simulations of the system. (a) Schematic illustration of the hybrid system based on single CdSe/ZnS QD coupled to a single GNR on ONF with a waist diameter of 320 nm.  b) Simulated Purcell enhancement factor (PF), emission enhancement factor into guided modes (EF) and degree of polarization (DOP) of the emitted photons into ONF guided modes by placing a dipole source (DS) near the surface of the GNR. The DS is kept 5 nm away from the surface of GNR (considering the size of a thick-shell QD) and its position is varied along the axis of GNR. $d$ denotes the axial position from the center of the GNR. (c) Simulated scattering field profile for z-polarization around the GNR.}
\label{Fig1}
\end{figure}

\section{Numerical Simulations}
A schematic diagram of the hybrid quantum system is shown in Fig. \ref{Fig1}(a). The hybrid quantum system is formed by combining a single QD and GNR with an ONF. Using the finite-difference time-domain (FDTD) method, we simulate the field profile, Purcell enhancement factor (PF), emission enhancement factor into guided modes (EF), and DOP of the emitted photons into ONF guided modes by placing a single dipole source (DS) close to the surface of the GNR. The PF is determined by $\Gamma/\Gamma_0$, where $\Gamma$ ($\Gamma_0$) is the total decay rate in the presence (absence) of GNR and ONF. The EF is determined by $I^{c}/I^{un}$, where $I^{c}$ and $I^{un}$ are the intensities of the emitted light coupled to the ONF in the presence and absence of the GNR, respectively. The DOP was determined by $(T_y+T_z-T_x)/(T_y+T_z+T_x)$, where $T_i$ ($i= x, y, z$) is the proportion of the emitted light coupled to the ONF for a dipole polarized along the $i$-axis.
The ONF is a silica cylinder of 320 nm in diameter, and the GNR is a gold rod with hemispherical end caps, with a length of 75 nm and a diameter of 30 nm. The GNR is placed on the ONF with its axis parallel to the ONF axis, and the DS is placed 5 nm away from surface of the GNR to mimic the experimental conditions. The DS wavelength is chosen from 550 to 750 nm. We perform the simulations for three polarization orientations (x,y,z) of the DS. The PF and EF are much strongly enhanced for the z-polarization compared to x- and y-polarization. The PF and EF values are estimated assuming random orientation of the dipole by averaging over the three polarizations. Figure \ref{Fig1}(b) shows the PF (black squares), EF (red dots) and DOP (blue stars) values for various dipole source positions ($d$) from the centre of the GNR. One can readily see the PF (EF) is highest at $d=0$ nm with a value around 60 (17) and rapidly decreases as $d$ increases. They reach a minimum at around $d=30$ nm and slightly increases after that. The DOP shows a much flat trend with a value in the range 90-99\%.

In order to understand the field enhancement due to the LSPR, we simulated the scattering field profile around the GNR. The simulated field profile for z-polarization is shown in Fig \ref{Fig1}(c). One can see that the field enhancement is maximum at the edges of the GNR as well as at its center. One can readily see that the field enhancement decreases rapidly as $d$ increases. It reaches minimum and again slightly increases due to the edge enhancement. This trend clearly explains the simulation results shown in Fig. \ref{Fig1}(b).


\section{Experimental Methods}

Figure \ref{Fig2}(a) shows the schematic diagram of the experiment. The experimental setup is based on a hybrid system of a single QD and a GNR deposited on an ONF. ONFs with optical transmission of $>$99\% were fabricated using a commercially available machine (Taper Fiber Expert, Deltafiber Ltd., Japan) by adiabatically tapering SMFs (SM 600, Fibercore) using a heat and pull technique. ONFs used for the experiment had a waist diameter of 310$\pm$10 nm and a uniform waist length of 2.5 mm. The ONF diameter was chosen to maximize the channeling of the single QD emission into the ONF guided modes \cite{famsan1,Yalla1}. The QDs used for the experiment were thick-shell CdSe/ZnS QDs dispersed in toluene colloidal solution \cite{Shafi1}. At room temperature, the QDs emit at a wavelength of 640 nm. The GNRs (E12-25-650, Nanopartz) used for the experiment were also dispersed in a toluene colloidal solution. The quoted diameter and length of the GNR were 25 and 71 nm, respectively. The quoted center wavelength of the LSPR of GNRs was 650 nm with a spectral width of around 100 nm FWHM.

The procedure for the sample preparation is as follows. The GNRs and QDs were deposited together on the ONF using a computer-controlled sub-pico-liter needle-dispenser (SPLD) system \cite{YallaOE} installed on an inverted microscope (Nikon, Eclipse Ti-U)). The dispenser consists of a tapered glass tube containing GNR and QD mixed in a toluene colloidal solution and a tungsten needle with a tip diameter of 5 $\mu$m. Once the computer-controlled needle tip passes through the glass tube, it carries a small amount of GNR-QD solution at its edge and a tiny amount of this solution is deposited on the ONF surface by bringing the tip into contact with the ONF. A laser was introduced into the ONF to monitor the deposition process. To get deterministic single GNR-QD on the ONF, we first optimized the concentration of GNR to be $10^{10}$ rods/cm$^{3}$ and that of QD to be $10^{13}$ dots/cm$^{3}$. The success probability of deposition of a single GNR-QD on the ONF for each trail was about 60\%.

Figure \ref{Fig2}(a) shows the measurement scheme of the experiment. Note that all experiments were performed at room temperature. The emission characteristics of the both coupled (GNR+QD) and uncoupled (only QD) single QDs were investigated via the guided mode of the ONF and eventually through a SMF. The single QDs were excited using a CW or a pulsed frequency-doubled YAG laser at a wavelength of 532 nm. The pulse width of the pulsed-laser was 20 ps. The excitation laser was linearly polarized with a polarization perpendicular to the ONF axis and laser is focused to a spot size of 1 $\mu$m using a microscope objective lens. 

The photo-luminescence (PL) from the single QDs is measured through the guided mode of the optical fiber on both sides of the ONF. The scattered excitation laser light is filtered using a 560 nm long-pass color glass filter (CF; (O56, HOYA)). Photons from one side of the fiber are used for spectral measurements using an optical multichannel spectrum analyzer (OMA). The same side is also used for emission polarization measurements. A fiber in-line polarization controller (not shown in Fig. \ref{Fig2}(a)) and a linear polarizer (P) are introduced in the path to the avalanche photodiode (APD) for the polarization measurements. It should be noted that for the LSPR spectrum measurements, the GNR on the ONF was irradiated with a white light source and the scattered light was measured through the ONF without CF.

Photons channelled into the other side of the fiber are used to measure the temporal characteristics. A Hanbury Brown-Twiss (HBT) scheme is employed for photon correlation measurements, in which photons emitted from the QD are split using a 50:50 non-polarizing beam splitter and detected by two fiber-coupled APDs. The arrival times of the photons are recorded using a time-correlated single photon counter (Picoharp 300, Pico Quant GmbH) for deriving the photon correlations. PL decay curves were obtained by measuring correlations between excitation sync pulse and arrival times of emitted photons.

\section {Results}

\begin{figure}[!ht]

\centering

 \includegraphics[width=\columnwidth]{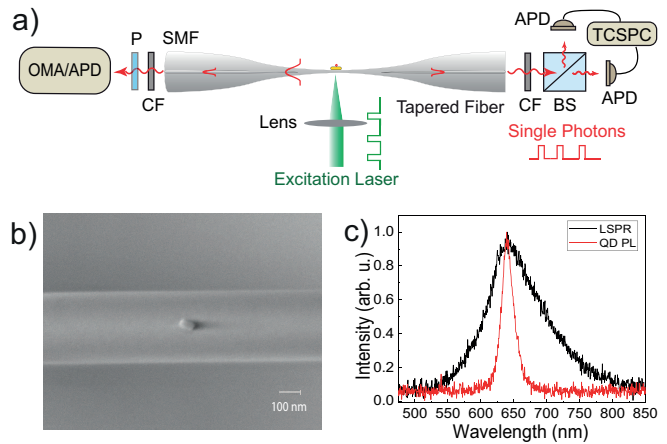}
 
\caption{Experimental Setup. (a) Schematic diagram of the hybrid system. A single CdSe/ZnS QD on a single GNR is interfaced with the ONF. The single QD is excited using a cw or a pulsed laser and the emission from the single QD is coupled to the guided mode of the ONF and eventually to a single mode optical fiber (SMF). The temporal, spectral and polarization characteristics of the fiber-guided photons are measured at either end of the fiber. The abbreviations denote, CF: color glass filter, BS: beam splitter, APD: avalanche photo-diode (single photon counting module), TCSPC: time-correlated single photon counter, P: polarizer, and OMA: optical multi-channel spectrum analyzer. (b) A typical SEM image of single GNR-QD deposited on ONF. (c) The black and red traces show the spectrum of light scattered by GNR indicating/revealing the LSPR of GNR and the PL spectrum of single QD, respectively.}
\label{Fig2}
\end{figure}

Figure \ref{Fig2}(b) shows a typical scanning electron microscope (SEM) image of the GNR-QD on the ONF. One can clearly see a single GNR on the surface of the ONF aligned along the ONF axis. From the SEM images, we estimated the size distribution of the GNR to be 71$\pm4$ in length and 31$\pm2$ nm in diameter. Note that the single QD is less than 10 nm in size and is not visible in the SEM image.

The black trace in Fig. \ref{Fig2}(c) shows the spectrum of light scattered by the GNR. One can clearly see a peak around 640 nm revealing the LSPR of the GNR. The FWHM of the LSPR spectrum is around 106 nm. The red trace in Fig. \ref{Fig2}(c) shows the PL-spectrum of the single QD on the ONF. The center wavelength of the PL-spectrum is 640 nm that matches to the LSPR peak and it has a FWHM of 19.6 nm that is well within the LSPR. The measured distribution of central wavelengths of QD and LSPR are $640\pm3$  and $640\pm5$ nm, respectively.

\begin{figure}[!ht]

 \includegraphics[width=\columnwidth]{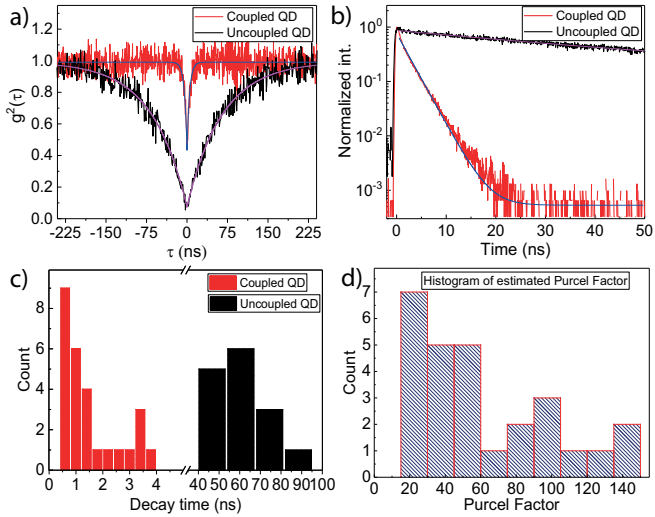}
 
\caption{Temporal characteristics of the PL-emission. (a) Normalized photon correlation $g^2(\tau)$ for coupled (red trace) and uncoupled (black trace) single QD, respectively. The solid curves show the exponential fit for the $g^2(\tau)$. (b) PL decay curves for coupled (red trace) and uncoupled (black trace) single QD, respectively. The solid curves show the fits using single exponential functions. (c) Histogram of decay times of coupled (red blocks) and uncoupled (black blocks) single QD, respectively. (d) Histogram of estimated Purcell factor (PF) for coupled single QDs.}
\label{Fig3}
\end{figure}

Figure \ref{Fig3} summaries the temporal characteristics of emission from coupled and uncoupled QDs on an ONF. We measured the photon correlations $g^2(\tau)$ of the PL-emission into the ONF to confirm that the emission is from a single QD. The red and black traces in Fig. \ref{Fig3}(a) show the normalized photon correlation $g^2(\tau)$ of the PL-emission from coupled and uncoupled QDs, respectively, while excited with the 532 nm cw-laser at an excitation intensity of 6 W/cm$^2$. The solid curves in Fig. \ref{Fig3}(a) are the exponential fit for $g^2(\tau)$ \cite{YallaOE}. One can clearly see the anti-bunching behaviour for both coupled and uncoupled QDs indicating that the emission is indeed from single QDs. The $g^2(0)$ values for the coupled and uncoupled cases are 0.43$\pm$0.03 and 0.07$\pm$0.02, respectively. The rise-times of the anti-bunching signals for the coupled and uncoupled QDs are 2.5$\pm$0.2 and 55$\pm$2 ns , respectively. This indicates the strong Purcell enhancement for the coupled QD. However, the rise-time of the anti-bunching signal also depends on the intensity of excitation.

To quantify the PF, we measured the PL decay time of the emission. The red and black traces in Fig. \ref{Fig3}(b) show the PL decay curves for coupled and uncoupled QDs, respectively. From the single exponential fits to the decay curves, we obtain the decay times of 2.6$\pm$0.1 ns and 61$\pm$2 ns for coupled and uncoupled QDs, respectively. The histograms of decay times for coupled and uncoupled QDs for different trials are shown as red and black blocks in Fig. \ref{Fig3}(c), respectively. From Fig. \ref{Fig3}(c), we estimate an average decay time of uncoupled QDs to be 55 ns. Using this average value, we estimate the PF for the coupled QDs. The histogram of PF for coupled QDs is plotted in Fig. \ref{Fig3}(d). From Fig. \ref{Fig3}(d), we infer that for majority of trials the PF lies between 20-60.

\begin{figure}[!ht]

 \includegraphics[width=\columnwidth]{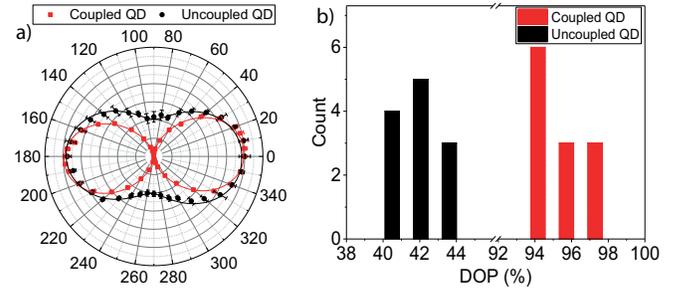}

\caption{Emission polarization characteristics of SPS. (a) Emission polarization characteristics of the SPS are indicated in the polar plot, the red data points are for the coupled and black data points are for uncoupled single QD. The solid curves are the cosine-squared fits. (b) Histograms of the degree of polarization (DOP) for coupled (red) and uncoupled (black) single QDs.}
\label{Fig4}
\end{figure}

Next, we present the polarization properties of the PL-emission from coupled and uncoupled QDs on the ONF. The QDs are excited with a polarization perpendicular to the ONF axis and the photon counts through the ONF are measured as a function of the polarizer angle. The red and black data points in Fig. \ref{Fig4}(a) show the polar plots for the polarization resolved normalized photon counts for coupled and uncoupled QDs, respectively. One can clearly see that for the coupled QD, the photon counts are strongly suppressed at an angle of 90$^{\circ}$ indicating high DOP. The solid curves are the cosine-squared fits. The DOP is estimated from the maximum ($N_{max}$) and minimum ($N_{min}$) photon counts, as $DOP=(N_{max}-N_{min})/(N_{max}+N_{min})$. The red and black blocks in Fig. \ref{Fig4}(b) show the histograms of the DOPs for the coupled and uncoupled QDs, respectively. One can see that the DOPs for uncoupled QDs are in the range of 40-44\%, whereas the DOPs for the coupled QDs are in the range of 94-97\%. This clearly indicates that the DOPs of single QDs can be strongly enhanced due to coupling to localized plasmon field of GNR \cite{Blinking2022}.

\begin{figure}[!ht]

 \includegraphics[width=\columnwidth]{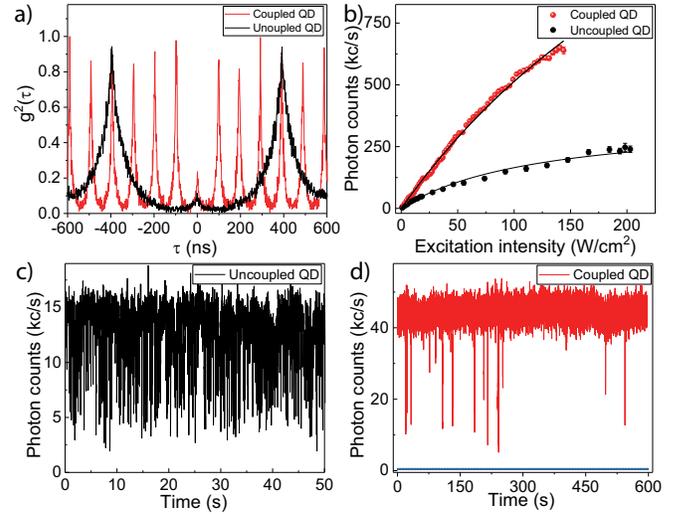}
 
\caption{Single photon characteristics of SPS. (a) The red trace shows the $g^{2}(\tau)$ of the coupled single QD measured by exciting the QD using a pulsed laser at 10 MHz. The black trace shows the $g^{2}(\tau)$ of the uncoupled single QD measured by pulsed excitation at 2.5 MHz (6 W/cm$^{2}$). (b) The red and blue data points show the photon count rate measured through one side of the fiber for coupled and uncoupled single QD respectively. The solid curves are the fits using the saturation model given in Eq.\ref{eq1}. (e) and (f) show the photon count rates for coupled and uncoupled single QDs, respectively. Both are measured with an excitation intensity of 6 W/cm$^{2}$}
\label{Fig5}
\end{figure}

Next, we evaluate the system performance as an efficient SPS. The purity of the fiber in-line SPS is estimated by measuring the photon correlations in a pulsed-excitation scheme. The red trace in Fig. \ref{Fig5}(a) shows the $g^{2}(\tau)$ of the coupled single QD measured by exciting the QD using a pulsed laser at 10 MHz repetition rate. The black trace in Fig. \ref{Fig5}(a) shows the $g^{2}(\tau)$ of the uncoupled single QD measured by pulsed excitation at a repetition rate of 2.5 MHz. One can clearly see that for both coupled and uncoupled QDs the central peak at $\tau$=0, is strongly suppressed indicating high purity single photon characteristics. The single photon purity for the coupled and uncoupled QDs are estimated to be $g^{2}(0)$=0.20$\pm$0.04 and 0.07$\pm$0.02, respectively. It should be noted that the reduced decay time for coupled QD enables excitation with a high repetition rate.

The brightness of the SPS is estimated by measuring the intensity dependence of the photon count rates as shown in Fig. \ref{Fig5}(b). The red and black data points in Fig. \ref{Fig5}(b) show the photon count rates for coupled and uncoupled QDs, respectively. It may be seen that the photon count rate increases with excitation intensity and shows a saturation behavior for higher excitation intensities. The observations are fitted using a saturation model for two-level system given as
\begin{equation}
{N(F)}= {N}_{max}[\frac{I}{I+I_{sat}}]
\label{eq1}
\end{equation}
where ${N}_{max}$ is the saturated photon count rate, ${I}$ and ${I}_{sat}$ are the excitation intensity and the saturation intensity, respectively. The solid curves in Fig. \ref{Fig5}(b) show the fits. From the fit for the coupled QD (uncoupled QD), we obtained a maximum count rate of single photons at one side of the fiber to be ${N}_{max}$= $2.50\pm0.13$ Mc/s ($0.37\pm0.17$ Mc/s) and a saturation intensity of ${I}_{sat}=387\pm25$ W/cm$^2$ ($122\pm9$ W/cm$^2$). 

The detection efficiency ($\alpha$) of the optical setup is estimated to be 41\%, which includes fiber coupling efficiency from ONF to APD (83\%), color glass filter transmission (83\%) and detection efficiency of APD at 650 nm (60\%). We estimated the brightness ($\Gamma_{SP}$) of the SPS (fiber-coupled photon count rate) as $\Gamma_{SP}$ = $2N_{max}/\alpha$. The factor two accounts for the coupling of single photons into both side of the ONF. The $\Gamma_{SP}$ for the coupled and uncoupled QDs are $12.2\pm0.6$ MHz and $1.8\pm0.8$ MHz, respectively.

We have also found that the stability of the photon count rate was improved for the coupled QDs. Typical photon count rates for uncoupled and coupled QDs are shown in Figs. \ref{Fig5}(c) and (d), respectively. The time axis resolution is 17 ms for both the measurements of 50 s (uncoupled QD) and 600 s (coupled QD). It can be seen that for uncoupled QD well-known strong blinking behaviour is observed \cite{Blinking2022,Shafi1}, whereas for the coupled QD such blinking behavior is strongly suppressed. This further improves the quality of the SPS.  

\section{Discussions}

A key experimental challenge in this work was to interface single GNR-coupled single QDs with ONF. Single QD deposition on ONF has been well-established using SPLD system \cite{YallaOE,Yalla1,Yalla3,Shafi1,ShafiSPS}. However, coupling single QDs to GNR on ONF was challenging. We have tried first depositing single QD and then GNR or vice-versa on the ONF. The success probability was very poor and repeated trials led to higher transmission loss of ONF. The procedure presented here works well with a high success probability of 60\% and typical transmission drop for each deposition was 1.4\%. One may further use sophisticated nano-manipulation techniques \cite{ONFNDarno,ChandraAFM} for accurate and deterministic positioning.

Another important observation is the orientation of the GNR axis on ONF. From the SEM images, we have found that the single GNRs are always aligned almost parallel to ONF axis. This may be due to the surface forces \cite{QuentinGNR}. One may use a composite technique \cite{ONFNDarno,ChandraAFM} or other nano-manipulation techniques to control the orientation of the GNR axis to further tune the coupling characteristics.

From the PF estimation in Fig. \ref{Fig3}(d), it is inferred that for majority of trials the PF is in the range of 20-60. This matches well with the theoretical simulations shown in Fig. \ref{Fig1}(b). This suggests that the assumption of random polarization orientation and position of QD on GNR are reasonable. Much higher PFs in the range of 80-160 observed in the experiment can be understood from specific polarization orientation or QDs deposited on the edges of GNRs (see Fig. \ref{Fig1}(c)). Such Purcell enhancement is much higher than that reported for ONF cavities \cite{Yalla3,Schell,Kali3, AokiCavity}. It is also broadband covering the entire PL-spectrum, useful for room-temperature based SPS. 

Apart from strong Purcell enhancement, another dramatic effect is the strong enhancement of DOP. As shown in Fig. \ref{Fig4}(b), the DOPs can be as high as 94-97\%. This also agrees well with the theoretical simulations shown in Fig. \ref{Fig1}(b). The single photons coupled to the fiber guided modes are strongly polarized along the transverse axis containing the QD and the GNR. Such high DOP is also essential for room temperature based SPS.

Regarding the purity of the SPS, the $g^{2}(0)$ values for coupled QDs are higher than that for uncoupled QDs (see Figs. \ref{Fig3}(a) and \ref{Fig5}(a)). We suspect that this may be related to the timing resolution of the detection system. The timing resolution of the detection system is 290 ps and the decay time of the coupled QD is 2.6 ns. Therefore, the timing resolution may induce some ambiguities in $g^{2}(0)$ values and PL decay times. We have found that for much shorter PL decay times $g^{2}(0)$ values show an increasing trend. For shorter PL decay times, the experimental data should be corrected for the effect of timing resolution.

From the $\Gamma_{SP}$ of the coupled and uncoupled QDs (discussed in Fig. \ref{Fig5}(b)), we estimate an EF of 6.8 for the coupled QD. From the PL decay times shown in Fig. \ref{Fig3}(b), the PF for the coupled QD is 23. These values are in good correspondence with the theoretical simulation shown in Fig. \ref{Fig1}(b). However, it should be noted that the coupling efficiency into ONF guided modes is reduced for the coupled QDs. From the $\Gamma_{SP}$ and PL decay times, we estimate a coupling efficiency of 3.2\% and 11\% for the coupled and uncoupled QDs, assuming 100\% quantum efficiency. The estimated coupling efficiencies also are in agreement with theoretical simulations (not shown here) assuming random polarization orientation of the dipole. The coupling efficiency of the coupled QD may be further improved by controlling the orientation of GNR axis or by combining it with a moderate finesse ONF cavity \cite{FamCavity,Yalla3,Schell}.

The blinking in the photon count rate (Fig. \ref{Fig5}c) of single QDs reveals that the emission switches between the exciton state and charged exciton (trion) state of the QD \cite{Shafi1,Louyer,Javaux}. The strong blinking behaviour in the photon count rate of uncoupled QD is due to the reduced quantum efficiency for the trion \cite{Shafi1,Gomez,Park2}. In contrast, the photon count rate of coupled QD (Fig. \ref{Fig5}d) shows strongly suppressed blinking, indicating a high radiative recombination rate of QDs in the presence of GNR. The present observations of the strongly suppressed blinking are in line with the previous report of CdSe/ZnS QDs in the proximity of gold nanorod/nanoantenna \cite{Blinking2010,Blinking2021,Blinking2022}.

\section{Conclusion}

In conclusion, we have demonstrated a fiber in-line SPS based on plasmon enhanced emission into ONF guided modes. We have shown that emission properties of single QDs can be strongly enhanced in the presence of GNRs leading to a bright and strongly polarized single photon emission efficiently coupled to the guided modes of the ONF and eventually to a SMF. In this hybrid system by combining GNR and ONF, one can take the advantages of both state-of-the-art field confinement and low-loss propagation of single photons in fiber guided modes. The localized surface plasmon field of GNR significantly improves the emission properties of single QDs (brightness, DOP, suppressed blinking) and the quality of SPS. Coupling the single photons into ONF guided modes enables automatic and alignment-free coupling to a SMF leading to a fiber-coupled SPS. The device can be integrated into the fiber networks paving the way for potential applications in quantum networks.  

\section{Acknowledgement}

We acknowledge Kohzo Hakuta and Makoto Morinaga for fruitful discussions. We thank Mark Sadgroove for the SEM measurements. We acknowledge the contributions of Akiharu Miyanaga, Kazunori Iida and Emi Tsutsumi, in the preparation of the thick shell quantum dot samples. KMS and KPN acknowledge support from Deltafibers Ltd. KPN acknowledges support from a grant-in-aid for scientific research (Grant No. 21H01011) from the Japan Society for the Promotion of Science (JSPS). RY acknowledges support from the Institute of Eminence (IoE) grant at the University of Hyderabad, India (File No. RC2-21-019).

\textit{Note added.}—After completion of this work, we have become aware of a related demonstration of plasmon-enhanced coupling of single photons into optical fiber \cite{MarkPlasmon}.

{}

\begin{thebibliography}{99}



\bibitem{Kimble} 
H. J. Kimble, The quantum internet, Nature \textbf{453}, 1023 (2008).

\bibitem{OBrien} 
J. O'Brien, A. Furusawa, and J. Vuckovic, Photonic quantum technologies, Nature Photon \textbf{3}, 687 (2009).

\bibitem{Arakawa}
K. Takemoto, Y. Nambu, T. Miyazawa, Y. Sakuma, T. Yamamoto, S. Yorozu and Y. Arakawa , Quantum key distribution over 120 km using ultrahigh purity single-photon source and superconducting single-photon detectors, Sci. Rep. \textbf{5}, 14383 (2015).

\bibitem{Aharonovich} I. Aharonovich, D. Englund, and M. Toth, Solid-state single-photon emitters, Nat. Photonics \textbf{10}, 631 (2016).

\bibitem{Senellart} P. Senellart, G. Solomon, and A. White, High-performance semiconductor quantum-dot single-photon sources, Nat. Nanotechnol. \textbf{12}, 1026 (2017).

\bibitem{Bennett}C. H. Bennett, and G. Brassard,  Quantum cryptography: Public key distribution and coin tossing,  International Conference on Computers, Systems and Signal Processing, Bangalore, India, Dec 9-12, (1984).

\bibitem{Nielsen}M. A. Nielsen and I. L. Chuang, Quantum computation and quantum communication (Cambridge University Press, Cambridge, 2000).

\bibitem{Nicolas} N. Gisin, G. Ribordy, W. Tittel, and H. Zbinden, Quantum cryptography, Rev. Mod. Phys. \textbf{74}, 145 (2002)

\bibitem{Michler} P. Michler, A. Kiraz, C. Becher, W. Schoenfeld, P. Petroff, L. Zhang, E. Hu, and A. Imamoglu, A quantum dot single-photon turnstile device, Science \textbf{290}, 2282 (2000).

\bibitem{Park} Y.-S. Park, J. Lim, and V. I. Klimov, Asymmetrically strained quantum dots with non-fluctuating single-dot emission spectra and subthermal room-temperature linewidths, Nat. Mater. \textbf{18}, 249 (2019).


\bibitem{Lin} X. Lin, X. Dai, C. Pu, Y. Deng, Y. Niu, L. Tong, W. Fang, Y. Jin, and X. Peng, Electrically-driven single-photon sources based on colloidal quantum dots with near-optimal anti-bunching at room temperature, Nat. Commun. \textbf{8}, 1132 (2017).

\bibitem{Senellart2} S.E. Thomas, M. Billard, N. Coste, S.C. Wein, Priya, H. Ollivier, O. Krebs, L. Tazaïrt, A. Harouri, A. Lemaitre, I. Sagnes, C. Anton, L. Lanco, N. Somaschi, J.C. Loredo, and P. Senellart, Bright polarized single-photon source based on a linear dipole, Phys. Rev. Lett. \textbf{126}, 233601 (2021).

\bibitem{Sipahigil} A. Sipahigil, R. E. Evans, D. D. Sukachev, M. J. Burek, J. Borregaard, M. K. Bhaskar, C. T. Nguyen, J. L. Pacheco, H. A. Atikian, C. Meuwly, R. M. Camacho, F. Jelezko, E. Bielejec, H. Park, M. Loncar, and M. D. Lukin, An integrated diamond nanophotonics platform for quantum-optical networks, Science \textbf{354}, 847 (2016).

\bibitem{Daveau} R. S. Daveau, K. C. Balram, T. Pregnolato, J. Liu, E. H. Lee, J. D. Song, V. Verma, R. Mirin, S. W. Nam, L. Midolo, S. Stobbe, K. Srinivasan, and P. Lodahl, Efficient fiber-coupled single-photon source based on quantum dots in a photonic-crystal waveguide, Optica \textbf{4}, 178 (2017).



\bibitem{Waks} C.-M. Lee, M. A. Buyukkaya,  S. Aghaeimeibodi,  A. Karasahin, C. J. K. Richardson, and  E. Waks, A fiber-integrated nanobeam single photon source emitting at telecom wavelengths, Appl. Phys. Lett. \textbf{114}, 171101 (2019).

\bibitem{Mulvaney} C. Sönnichsen, T. Franzl, T. Wilk, G. von Plessen, J. Feldmann, O. Wilson, and P. Mulvaney, Drastic reduction of plasmon damping in gold nanorods, Phys. Rev. Lett. \textbf{88}, 077402 (2002).

\bibitem{Esteban}R. Esteban, T. V. Teperik, and J. J. Greffet, Optical patch antennas for single photon emission using surface plasmon resonances, Phys. Rev. Lett. \textbf{104}, 026802 (2010)

\bibitem{Gleb} G. M. Akselrod, C. Argyropoulos, T. B.Hoang, C. Ciracì, C. Fang, J.Huang, D. R. Smith, M. H. Mikkelsen, Probing the mechanisms of large Purcell enhancement in plasmonic nanoantennas, Nat. Photonics \textbf{8}, 835 (2014).

\bibitem{TongGNR} P. Wang, Y. Wang, Z. Yang, X. Guo, X. Lin, X.-C. Yu, Y.-F. Xiao, W. Fang, L. Zhang, G. Lu, Q. Gong, and L. Tong, Single-band 2-nm-line-width plasmon resonance in a strongly coupled Au nanorod,
Nano Lett. \textbf{15}, 7581 (2015). 

\bibitem{Hoang} T. B. Hoang, G. M. Akselrod, and M. H. Mikkelsen, Ultrafast room-temperature single photon emission from quantum
dots coupled to plasmonic nanocavities, Nano Lett. \textbf{16}, 270 (2016).

\bibitem{Blinking2022} Y. He, J. Chen, R. Liu, Y. Weng, C. Zhang, Y. Kuang, X. Wang, L. Guo, and X. Ran, Suppressed Blinking and Polarization-dependent emission enhancement of single ZnCdSe/ZnS dot coupled with Au nanorods, ACS Appl. Mater. Interfaces \textbf{14}, 12901 (2022).
\bibitem{LTong1} L. Tong, R. R. Gattass, J. B. Ashcom, S. He, J. Lou, M. Shen, I. Maxwell, and E. Mazur, Subwavelength-diameter silica wires for low-loss optical wave guiding, Nature \textbf{426}, 816 (2003).
\bibitem{famsan1} F. L. Kien, S. D. Gupta, V. I. Balykin, and K. Hakuta, Spontaneous emission of a cesium atom near a nanofiber: Efficient coupling of light to guided modes, Phys. Rev. A \textbf{72}, 032509 (2005).

\bibitem{Sile1}T. Nieddu, V. Gokhroo, and S. Nic Chormaic, Optical nanofibres and neutral atoms, J. Opt. \textbf{18}, 053001 (2016).

\bibitem{Rolston}P. Solano, J. A. Grover, J. E. Homan, S. Ravets, F. K. Fatemi, L. A. Orozco, and S. L. Rolston, Optical nanofibers: A new platform for quantum optics, Adv. At. Mol. Opt. Phys. \textbf{66}, 439 (2017).

\bibitem{Kali1} K. P. Nayak, M. Sadgrove, R. Yalla, F. L. Kien and K. Hakuta, Nanofiber quantum photonics, J. Opt. \textbf{20}, 073001 (2018).

\bibitem{LTong2} L. Shao, H. Wu, W. Fang, and L. Tong, Twin-nanofiber structure for a highly efficient single-photon collection, Opt. Express \textbf{30}, 9147 (2022).

\bibitem{Kali2} K. P. Nayak, P. N. Melentiev, M. Morinaga, F. L. Kien, V. I. Balykin, and K. Hakuta, Optical nanofiber as an efficient tool for manipulating and probing atomic fluorescence, Opt. Express \textbf{15}, 5431 (2007).

\bibitem{Yalla1} R. Yalla, F. L.Kien, M. Morinaga, and K. Hakuta, Efficient channeling of fluorescence photons from single quantum dots into guided modes of optical nanofiber, Phys. Rev. Lett. \textbf{109}, 063602 (2012).

\bibitem{ONFNDarno} L. Liebermeister, F. Petersen, A. V. Münchow, D. Burchardt, J. Hermelbracht, T. Tashima, A. W. Schell, O. Benson, T. Meinhardt, A. Krueger, A. Stiebeiner, A. Rauschenbeutel, H. Weinfurter, and M. Weber, Tapered fiber coupling of single photons emitted by a deterministically positioned single nitrogen vacancy center, Appl. Phys. Lett. \textbf{104}, 031101 (2014).

\bibitem{FujiwaraQD} M. Fujiwara, K. Toubaru, T. Noda, H.-Q. Zhao, and S. Takeuchi, Highly efficient coupling of photons from nanoemitters into single-mode optical fibers, Nano Lett. \textbf{11}, 4362 (2011).



\bibitem{Skoff} S. M. Skoff, D. Papencordt, H. Schauffert, B. C. Bayer, and A. Rauschenbeutel, Optical-nanofiber-based interface for single molecules, Phys. Rev. A \textbf{97}, 043839 (2018).

\bibitem{Schell2}A. W. Schell, H. Takashima, T. T. Tran, I. Aharonovich, and S. Takeuchi, Coupling quantum emitters in 2D materials with tapered fibers, ACS Photonics \textbf{4}, 761 (2017).

\bibitem{aoki} S. Kato and T. Aoki, Strong coupling between a trapped single atom and an all-fiber cavity, Phys. Rev. Lett. \textbf{115,} 093603 (2015).

\bibitem{Kali3} K. P. Nayak, J. Wang, and J. Keloth, Real-time observation of single atoms trapped and interfaced to a nanofiber cavity, Phys. Rev. Lett. \textbf{123}, 213602 (2019).


\bibitem{ShafiSPS} K. M. Shafi, K. P. Nayak, A. Miyanaga, and K. Hakuta, Efficient fiber in‑line single photon source based on colloidal single quantum dots on an optical nanofiber, Applied Physics B \textbf{126}, 58 (2020).

\bibitem{YallaOE} R. Yalla, K. P. Nayak, and K. Hakuta, Fluorescence photon measurements from single quantum dots on an optical nanofiber, Opt. Express \textbf{20}, 2932 (2012).

\bibitem{Louyer} Y. Louyer, L. Biadala, P. Tamarat, and  B. Lounis, Spectroscopy of neutral and charged exciton states in single CdSe/ZnS nanocrystals, Appl. Phys. Lett. \textbf{96}, 203111 (2010).

\bibitem{Shafi1} K. M. Shafi, W. Luo, R. Yalla, K. Iida, E. Tsutsumi A. Miyanaga and K. Hakuta, Hybrid system of an optical nanofibre and a single quantum dot operated at cryogenic temperatures, Sci. Rep. \textbf{8}, 13494 (2018).

\bibitem{Javaux} C. Javaux, B. Mahler, B. Dubertret, A. Shabaev, A. V. Rodina, Al. L. Efros, D. R. Yakovlev, F. Liu, M. Bayer, G. Camps, L. Biadala, S. Buil, X. Quelin and J-P. Hermier, Thermal activation of non-radiative Auger recombination in charged colloidal nanocrystals, Nature Nanotech. \textbf{8}, 206 (2013).

\bibitem{Park2}  Y. S. Park, W.K. Bae, J. M. Pietryga and Klimov, V. I. Auger recombination of biexcitons and negative and positive trions in individual quantum dots, ACS Nano \textbf{8}, 7288 (2014).

\bibitem{Gomez} D. E. Gómez, J. Embden, P. Mulvaney, M. J. Fernée, and H. R. Dunlop, Exciton-trion transitions in single CdSe-CdS core-shell nanocrystals, ACS Nano \textbf{3}, 2281 (2009).

\bibitem{Blinking2010} X. Ma, H. Tan, T. Kipp, and A. Mews, Fluorescence enhancement, blinking Suppression, and gray states of individual semiconductor nanocrystals close to gold nanoparticles, Nano Lett. \textbf{10}, 4166 (2010).

\bibitem{Blinking2021} Q. Jiang, P. Roy,  J. B. Claude, J.Wenger, Single photon source from a nanoantenna-trapped single quantum dot, Nano Lett. \textbf{21}, 7030 (2021).

\bibitem{QuentinGNR} M. Joos, C. Ding, V. Loo, G. Blanquer, E. G., A. Bramati, V. Krachmalnicoff, and Q. Glorieux, Polarization control of linear dipole radiation using an optical nanofiber, Phys. Rev. Applied \textbf{9}, 064035 (2018)

\bibitem{ChandraAFM}R. Yalla, Y. Kojima, Y. Fukumoto, H. Suzuki, O. Ariyada, K. M. Shafi, K. P. Nayak and K. Hakuta, Integration of silicon-vacancy centers in nanodiamonds with an optical nanofiber, Appl. Phys. Lett. \textbf{120}, 241102 (2022).

\bibitem{FamCavity} F. L. Kien and K. Hakuta, Cavity-enhanced channeling of emission from an atom into a nanofiber, Phys. Rev. A \textbf{80}, 053826 (2009).

\bibitem{Yalla3} R. Yalla, M. Sadgrove, K. P. Nayak, and K. Hakuta, Cavity quantum electrodynamics on a nanofiber using a composite photonic crystal cavity, Phys. Rev. Lett. \textbf{113}, 143601 (2014).
\bibitem{Schell} A. W. Schell, H. Takashima, S. Kamioka, Y. Oe, M. Fujiwara, O. Benson and  S. Takeuchi, Highly efficient coupling of nanolight emitters to a ultra-wide tunable nanofibre cavity, Sci. Rep. \textbf{5}, 9619 (2016).

 
\bibitem{AokiCavity}S. K. Ruddell, K. E. Webb, M. Takahata, S. Kato, AND T. Aoki, Ultra-low-loss nanofiber Fabry–Perot cavities optimized for cavity quantum electrodynamics, Optics Letters, \textbf{45}, 4875 (2020).

\bibitem{MarkPlasmon}M. Sugawara, Y. Xuan, Y. Mitsumori, K. Edamatsu, M. Sadgrove, Plasmon-enhanced polarized single photon source directly coupled to an optical fiber, arXiv:2203.01591 (2022).




\end{thebibliography}
\end{document}